\documentclass[conference]{IEEEtran}
\IEEEoverridecommandlockouts
\usepackage{cite}
\usepackage{amsmath,amssymb,amsfonts}
\usepackage{algorithmic}
\usepackage{graphicx}
\usepackage{textcomp}
\usepackage{xcolor}
\usepackage{booktabs}
\usepackage{bm}
\usepackage{array}
\usepackage{longtable}
\usepackage[ruled]{algorithm}
\usepackage{subcaption}
\usepackage[flushleft]{threeparttable}

\newcommand{\argmin}{\mathop{\rm arg~min}\limits}

\usepackage{geometry}
\usepackage{enumitem}
\newtheorem{proposition}{Proposition}
\newtheorem{proof}{Proof}

\def\BibTeX{{\rm B\kern-.05em{\sc i\kern-.025em b}\kern-.08em
    T\kern-.1667em\lower.7ex\hbox{E}\kern-.125emX}}
\begin{document}

\title{A Real-World Implementation of Unbiased Lift-based Bidding System\\
 \thanks{We thank Shota Yasui for the insightful comments, Shin Goto, Yuki Toyoda, Naoya Yokoyama, and Nozomu Fujiwara for helping our experiments.}
}

\author{\IEEEauthorblockN{Daisuke Moriwaki}
\IEEEauthorblockA{\textit{AI Lab} \\ \textit{AI Division, CyberAgent, Inc.} \\
Tokyo, Japan \\
moriwaki\_daisuke@cyberagent.co.jp}
\and
\IEEEauthorblockN{Yuta Hayakawa}
\IEEEauthorblockA{\textit{AI Division, CyberAgent, Inc.} \\
Tokyo, Japan \\
hayakawa\_yuta@cyberagent.co.jp}
\and
\IEEEauthorblockN{Akira Matsui}
\IEEEauthorblockA{\textit{Department of Computer Science} \\
\textit{University of Southern California} \\
Los Angels, CA  \\
amatsui@usc.edu}
\and
\IEEEauthorblockN{Yuta Saito}\thanks{This work was done when YS was at Hanjuku-kaso Co Ltd, Tokyo, Japan.}
\IEEEauthorblockA{\textit{Department of Computer Science} \\
\textit{Cornell University} \\
Ithaca, NY \\
ys552@cornell.edu}
\and
\IEEEauthorblockN{Isshu Munemasa}
\IEEEauthorblockA{\textit{AI Division, CyberAgent, Inc.} \\
Tokyo, Japan \\
munemasa\_isshu@cyberagent.co.jp}
\and
\IEEEauthorblockN{Masashi Shibata}
\IEEEauthorblockA{\textit{AI Lab} \\ \textit{AI Division, CyberAgent, Inc.} \\
Tokyo, Japan \\
masashi\_shibatau@cyberagent.co.jp}
}
\IEEEoverridecommandlockouts
\maketitle
\IEEEpubidadjcol
\begin{abstract}
In display ad auctions of Real-Time Bidding (RTB), a typical Demand-Side Platform (DSP) bids based on the predicted probability of click and conversion right after an ad impression. Recent studies find such a strategy is suboptimal and propose a better bidding strategy named lift-based bidding. Lift-based bidding simply bids the price according to the lift effect of the ad impression and achieves maximization of target metrics such as sales. Despite its superiority, lift-based bidding has not yet been widely accepted in the avertising industry. For one reason, lift-based bidding is less profitable for DSP providers under the current billing rule. Second, the practical usefulness of lift-based bidding is not widely understood in the online advertising industry due to the lack of a comprehensive investigation of its impact. 

We here propose a practically-implementable lift-based bidding system that perfectly fits the current billing rules. We conduct extensive experiments using a real-world advertising campaign and examine the performance under various settings. We find that lift-based bidding, especially unbiased lift-based bidding is most profitable for both DSP providers and advertisers. Our ablation study highlights that lift-based bidding has a good property for currently dominant first price auctions. The results will motivate the online advertising industry to consider lift-based advertising.
\end{abstract}

\section{Introduction}
\label{sec: introduction}
Online display advertising has been essential for the recent business, which accounts for half of the US advertiser's expenditures~\cite{USProgrammaticDigital}. Ad deliverers, Demand-Side Platforms (DSPs) charge ad costs for advertisers through the objective billing rules named cost-per-click (CPC) and cost-per-action (CPA), by which advertisers pay DSPs a fixed cost for each click or conversion. To maximize click-charge and conversion-charge, most of the DSPs have been following the ``performance-based bidding'' strategy which determines bid price based on the probability of users taking the desired action (attributed action) after the ad delivery. 

Despite its industrial success, researchers have recognized a serious caveat in the bidding process~\cite{lewis2018incrementality,xu2016lift,moriwakiUnbiasedLiftbasedBidding2020}. The performance-based bidding strategy ignores the probability that a user will convert even without an ad. Such a strategy is suboptimal since it will not reach users with a high probability of \textit{changing} their actions by showing an ad. Moreover, it might dissuade end-users from conversion~\cite{agarwalSpatiotemporalModelsEstimating2009a,maUserFatigueOnline2016a,moriwaki_fatigue-aware_2020}.

Recently, not only scholars but also practitioners in the display advertising industry pay stronger attention to the lift effect (causal effect, incrementality) of advertising. Major DSP providers such as Criteo and Yahoo! advocate the importance of the causal effects of advertising~\cite{criteoIncrementalitySimpleQuestion,barajas_online_2021,beeswaxMeasuringIncrementalityDigital}. It is urgent for DSP providers to shift from click/conversion-maximization to lift-maximization as the client advertisers want real success in the advertising campaign rather than the maximization of attributed clicks and conversions.

Xu et al. (2016)~\cite{xu2016lift} is the seminal work that studies the lift-based bidding algorithm. They show that lift-based bidding is theoretically more efficient to increase sales than performance-based bidding and demonstrate it in the online experiment. 

They left two challenges for the real-world implementation of lift-based bidding. First, their lift-effect predictor does not correct for the bias inherited from the training data. Since the data used for training is the results of past advertising campaigns, the ad exposures are not randomized but biased by targeting strategy. Second, as they have clearly shown in the paper, lift-based bidding can not be implemented in the real-world. Under the current billing rule, DSPs are only rewarded when they earn attributed actions but not when they change users' actions, i.e. conversion lift.

Moriwaki et al.~\cite{moriwakiUnbiasedLiftbasedBidding2020} addressed the former challenge by introducing the unbiased lift-effect predictor in the bidding system and showed its superiority in a real-world online experiment. However, they only compare the lift-based bidder and the conventional production-ready bidder. They failed to demonstrate how their ``unbiased'' predictor affected the result. More importantly, same as \cite{xu2016lift}, the proposed system is not ready for release as a product because their system is not profitable for DSP providers under the current billing rule.

In this paper, we address the challenges left by the literature. First, we propose an implementable lift-based bidder by combining click-through rate (CTR) predictor and lift-effect predictor in the bidding system. Second, we deploy the unbiased lift-based bidder and compare them with three variants including one based on lift-based without debiasing~\cite{xu2016lift}, unbiased lift-based~\cite{moriwakiUnbiasedLiftbasedBidding2020}, and unbiased lift-based with clipping in the real-world online experiment.

We find that an unbiased lift-based bidding system achieved the best cost-per incremental action (CPIA), which is equivalent to the highest return on advertising (ROA) for the advertiser. At the same time, the proposed system achieved the highest CTR, which implies the profitability for DSP providers.

Furthermore, we conduct a detailed analysis of the result and find lift-based bidders' bid prices are closer to the clearing price than the performance-based bidder. This property helps DSPs save inventory costs (cost to buy ad impressions). Lift-based bidders have practical advantages in the first-price auction that is dominant in the current display advertising scenes.

In sum, our contribution is 
{
\setlength{\leftmargini}{10pt}  
\begin{itemize}
    \item to propose the lift-based bidding system that is implementable under the current billing rule,
    \item to conduct comprehensive online experiments in the real-world advertising campaign and show the competitiveness of unbiased lift-based bidding, and
    \item to show that lift-based bidders' bid price is closer to clearing price than the conventional bidder which is essential to achieve cost-efficiency in the first-price auctions
\end{itemize}.
}
The present work provides the online advertising community with insights into how lift-based bidding system work and encourage serious considerations on lift-based strategy.

\section{Background}\label{sec:background}
To further motivate our work, we present the difference between lift-based bidding and conventional performance-based bidding and explain the challenges that DSPs face. 
For readers' convenience, we summarize the technical terms in Table\ref{tab:glosarry}.
\begin{table}[htpb]  \small
    \centering
        \caption{Glossary of Terms}
    \begin{tabular}{p{25mm}p{40mm}}
    \toprule
    \multicolumn{1}{c}{Term} & \multicolumn{1}{c}{Definition} \\ \midrule
        Demand-Side Platform (DSP) & A server operated by ad-tech companies that participates ad auction to buy ad-slot. \\
        Cost-per-click (CPC) billing & Advertisers pay DSP providers fixed cost for attributed clicks.\\
        Cost-per-action (CPA) billing. & Advertisers pay DSP providers fixed costs for attributed conversions (e.g. visit).\\
        CPC/CPA charge & Cost borne by advertisers according to the CPC/CPA billing rule. \\
        Inventory cost & Cost borne by DSP providers to buy impression. The price is determined by auction. \\
        Incremental action/visit & Increase in conversions due to advertising.  \\
        Return on Advertising (ROA) & Increase in sales per advertising cost \\
        Cost-per incremental action (CPIA) & Total incremental actions divided by CPC/CPA charge\\
        \bottomrule
    \end{tabular}
    \label{tab:glosarry}
\end{table}
\subsection{Performance-based vs. Lift-based Bidding}
\begin{figure}[htb]
  \centering \includegraphics[width=0.95\linewidth]{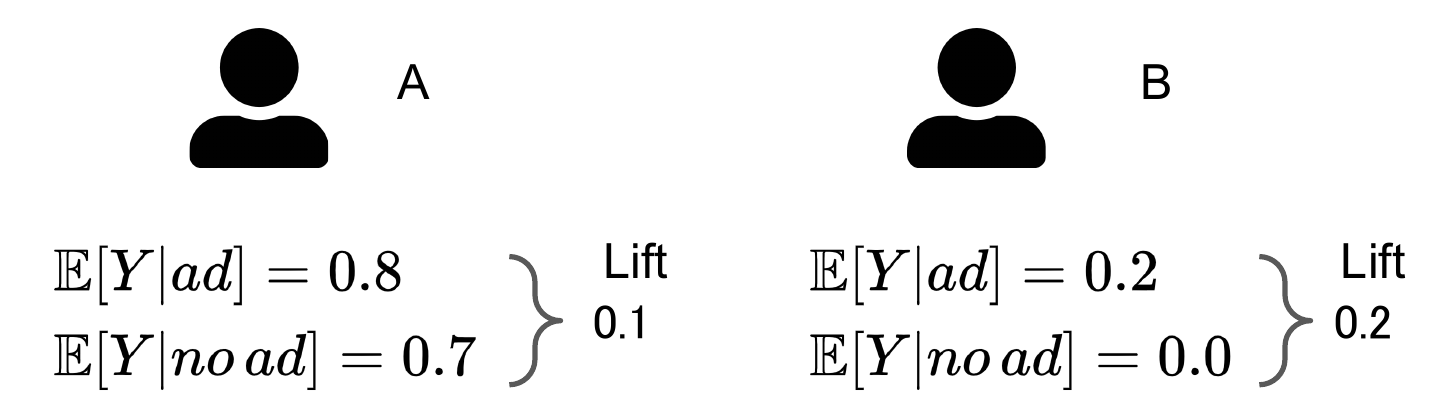}
  \caption{Schematic of the lift effect}  \label{fig:lift}
  \begin{minipage}{\columnwidth}%
    {\small {\it Note}: Customer A has a high probability of conversion without ad while Customer B is more responsive.}
  \end{minipage}%
  \end{figure}

Let $Y$ be a binary variable that takes one when the consumer purchase the target product(s) and takes zero when not. Then the conditional conversion rate given advertisement is $\mathbb{E}[Y|\rm{ad}]$ while that given no advertisement is $\mathbb{E}[Y|\rm{no \, ad}]$. In Fig.~\ref{fig:lift}, customer A will convert (purchase) at a probability of 0.8 when she is exposed to advertisement and 0.7 when not. Customer B will convert at a probability of 0.2 when he is exposed to the advertisement and 0.0 when not. 

Conventional bidders bid higher prices for Customer A because it only sees the conditional conversion rate (0.8 vs. 0.2) while advertisers value the ad for B from the viewpoint of {\it lift} (0.1 vs. 0.2). Performance-based billing does not give the reward for delivering ads to responsive customers (B) but for finding customers who are prone to convert regardless of an ad impression (A).
In other words, the conventional strategy that maximizes the number of attributed conversions, $\mathbb{E}[Y|{\rm ad}]$. A better objective to be maximized is the lift effect, $\mathbb{E}[Y|{\rm ad}] - \mathbb{E}[Y|{\rm no \, ad}]$.

\subsection{The Challenges of the Lift-based Bidding}
To maximize the lift effect, it is natural for DSP to determine the bid price based on the lift effect rather than the probability of attributed actions. However, the majority of DSPs follow a performance-based bidding strategy and pursue attributed clicks and conversions, which is not directly linked to the advertisers' goal.

This gap between DSPs' strategy and advertisers' goal stems from the gap in their target metrics. The advertisers want to maximize the return on advertising (ROA) which is, in turn, the minimization of cost-per incremental action (CPIA, cost paid by advertiser for unit of lift effect). On the other hand, DSPs are simply pursuing the number of clicks and conversions associated with the ad. 

Unfortunately, as proved in \cite{xu2016lift}, lift-based bidding is not profitable for DSP providers under the current billing rule. Lift-based bidding buys ad-slots with larger lift effects. However, they are only rewarded for attributed clicks (conversions) but not for lift-effect. In other words, lift-based bidding sacrifices DSP's profit to increase advertiser's sales.

Ideally, there should be a {\it lift-based billing rule} for advertisers who want to minimize CPIA, which resolves this conflict of interest. The problem is a lift-based billing rule is hard to be implemented in practice. 

First of all, the lift effect is counterfactual (i.e., not observable) so that it is always need to be estimated. There is no guarantee that the stakeholders all agree on one estimate. Second, conventional performance-based billing is simple and easy to understand while the understanding of the importance of lift-based advertising is limited. There is little incentive for the industry to change the de-facto standard. 

In sum, lift-based bidding inevitably needs modifications to be introduced in the advertising industry. Moreover, such modifications should be tested in real-world advertising campaigns and proved to be practically effective. 

\section{Related Work}\label{sec:related work}
\subsection{Real-Time Bidding (RTB) System}
RTB is a programmatic infrastructure where the publishers sell impressions of users to advertisers through online auctions.  In the RTB environment~\cite{yuan2013real,wang2015real,wang2016display}, DSPs participate in online auctions to purchase ad impressions. DSP charges advertisers for the cost based on the observed metrics, such as the number of clicks and conversions after ad impressions. Bidding the true value is well known to be a dominant strategy in the second price auction. However, the recent literature has realized that this simple strategy does not hold under the budget constraint~\cite{balseiro2015repeated}, and pacing strategy could be the optimal strategy~\cite{balseiro2017budget, balseiro2019learning} and for first-price auction~\cite{conitzer2018pacing}. Given that most of the SSPs transformed to a first-price auction mechanism, bid shading becomes essential for DSPs~\cite{panBidShadingWinRate2020}.
\subsection{Performance-based Bidding Strategy} \label{sec:relatedwork performance-based}
Most of the existing researches propose performance-based bidding. To implement this automatic system, the bidding system has adopted several implementations such as reinforcement learning~\cite{wu2018budget, cai2017real}, proportional-integral-derivative (PID) controller~\cite{yang2019bid,maehara2018optimal,zhang2016feedback}.  Since the performance of the bidding system has been evaluated by the reward to the DSPs, previous studies have proposed methods to predict the user responses to given an ad impression. 

In the online advertising industry, the goal of the advertisers is naively defined as the number of clicks and conversions after ad impressions ~\cite{lewis2014online,cai2017real}. The researchers have proposed a prediction model for click-through rate (CTR)~\cite{cheng2010personalized,zhu2010novel,ren2016user,qu2016product} or conversion rate (CVR)~\cite{rosales2012post,yeo2017predicting, qu2016product}.

\subsection{Lift-based Bidding Strategy and Impression Bias}
Recently, many researchers study predictions of a causal effect of policy interventions as {\it uplift modeling}~\cite{radcliffe2011real,jaskowski2012uplift,rzepakowski2012decision,zaniewicz2013support,diemertLargeScaleBenchmark2018,kawanaka2019,saito2019doubly,saito2020cost}. Attempts to incorporate the causal effect of the ad in the bidding strategy are proposed by several researchers including lift-based bidding~\cite{xu2016lift}, incrementality bidding~\cite{johnsonCostIncrementalAction2015}, and unbiased lift-based bidding~\cite{moriwakiUnbiasedLiftbasedBidding2020}.

 \cite{xu2016lift} first show that lift-based bidding is more efficient than performance-based bidding theoretically and empirically. A problem with this work is that they ignore the effect of the impression bias. A more appropriate approach needs to incorporate causal inference technique~\cite{bottou_counterfactual_2013,joachims_unbiased_2018}. \cite{moriwakiUnbiasedLiftbasedBidding2020} addresses the inherent bias in impression log data. Specifically, they propose a method to unbiasedly predict the lift-effect of an ad impression on a specific user from biased impression data. The method is easily implementable with the well-known machine learning libraries, while the previous debiasing method for the performance-based bidding strategy requires additional implementation cost~\cite{zhang2016bid}. In another strand, Bompaire et al. \cite{bompaire_causal_2021} propose a rigorous attribution model based on a causal model and successfully show the cost-reduction in real-world experiment. However, their work is different from ours since they assume no conversion when an ad is not delivered. Thus, the previous works have a limitation in their implementability in the real-world ad-tech industry because there is no incentive for DSPs to pursue lift-effect in the conventional billing rule. The present work attempts to solve the inconsistency between the prevailing billing rule and the advertisers' true goal. 

\section{Proposed Method}
\label{sec: proposed method}
We propose an unbiased lift-based bidding strategy under a performance-based billing rule. While we describe the case of CPC billing due to its popularity, it can be extended to the cost-per-action (CPA) version. We use the inverse propensity score (IPS) technique and provide proof for unbiasedness.

\subsection{Setup}
We consider a bidding strategy that maximizes the number of clicks and minimizes CPIA by maximizing lift-effect at the same time. To this end, the bid price should be based on both the lift effect of advertising and predicted CTR (pCTR). The combination of lift effect and CTR prediction is expected to contribute to both an advertiser's sales and DSP's profits. 
In particular, our algorithm calculates $bid_{t}$, bid price for $t$-th auction as:
\begin{align}\label{eq:bidding}
    bid_{t} &=& \phi ( \mathbf{x}_i, s(a) ) \cdot  CPC \cdot pCTR \cdot \alpha.
\end{align}
where $\phi ( \mathbf{x}_i, s(a) )$ denotes normalized predicted lift-effect when it delivers additional ad $a$ to user $i$ who has feature vector $\mathbf{x}_i$. In the performance based bidding $\phi$ becomes a normalized predicted CVR. The details are described below. $CPC$ is a fixed reward for each click, $pCTR$ is the predicted CTR of the ad slot. CTR prediction is a well-studied task (Sec.\ref{sec:relatedwork performance-based}). We simply use the predictor deployed in the production environment. $\alpha \in (0,1)$ is a budget pacing multiplier (Section~\ref{sec:automated bid adjustment}). In the first-price auction, $\alpha$ is also a bid shading parameter.

$\phi ( \mathbf{x}_i, s(a) )$ is the most important part for lift based bidding. Specifically, $\phi(\mathbf{x}_i, s(a))$ is calculated using the following equation:
\begin{eqnarray} \label{eq:phi}
    \phi(\mathbf{x}_i, s(a)) = \frac{\tau(s(a) \mid \mathbf{x}_i)}{\bar{\tau}},
\end{eqnarray}
where $\tau(s(a) \mid \mathbf{x}_i)$ is the predicted lift-effect of additional exposure to the ad $a$ for user $i$ characterized by a feature vector $\mathbf{x}_i$. 
By dividing by the mean lift effect $\bar{\tau} = \frac{1}{|\mathcal{I}||\mathcal{S}|}\sum\limits_{i, s(a)}\tau(s(a)|\mathbf{x}_i)$, $\phi(\cdot)$ is normalized so that $E[\phi(\cdot)] = 1$. This normalization stabilizes bid prices when combined with pCTR.

$s(a) \in \mathcal{S}$ represents the {\it ad exposure state} of an ad $a$ under consideration, and $\mathcal{S}$ is a set of possible states. We use the number of impressions of $a$ to $i$ as a scalar variable representing the ad exposure state of $a$ to $i$, and thus $\mathcal{S} = \{0, 1, \ldots \}$ 

To formally define the lift-effect $\tau$, we introduce the essential notation called \textit{potential outcome} in causal inference~\cite{imbens2015causal}. Let $y_i (s(a))$ denote user $i$'s potential outcome associated with the exposure state $s(a)$ of ad $a$. Each user $i$ has potential outcomes associated with every possible state, that is, $ \mathbf{y} = \{ y (s(a)) \ | \ \forall s(a) \in S \}$, however, only one of them is observable. The observed outcome for user $i$ is defined as $ y_i^{obs} =  y_i (s_i)$ where $s_i$ is a random variable representing an exposure state for user $i$. Note that the potential outcomes associated with every possible state other than the realized one is unobservable.

The lift-effect $\tau$ of showing ad $a$ for each user $i$ is sequentially defined as the difference between the expectation of the potential outcomes given the two consecutive ad exposure states (the number of impressions):
\begin{align}\label{eq:tau} \small \nonumber
\tau(s (a) & \ | \ \mathbf{x}_i) 
 = E[ y_i (s(a)) \ | \ \mathbf{x}_i ] -  \\ &E[ y_i (s(a) - 1) \ |\ \mathbf{x}_i] , \forall s(a) \in \mathcal{S} \backslash \{0\},
\end{align}
where $ E[ y_i (s(a)) |\mathbf{x}_i  ] $ is the expected potential outcome of $i$ when the number of impressions is $s(a)$. In contrast, $ E[ y_i (s(a) - 1) |\mathbf{x}_i] $ is the expected potential outcome when the number of impressions is $s(a)-1$. Thus, Eq.~(\ref{eq:tau}) is a reasonable definition for the lift-effect of showing an additional ad $a$ to a specific user $i$ who was exposed to the ad $s(a)-1$ times in the past. This formulation well captures wear-in and wear-out effects of additional impression~\cite{lewis_worn-out_2015,moriwaki_fatigue-aware_2020}. Note that $\tau (0 \ | \ \mathbf{x}_i)=0$ by definition while $E[ y_i (0) |\mathbf{x}_i] \geq 0$ (organic conversion).

To predict $\tau$, we train predictors of outcomes for every possible state separately and combine their predictions as follows. 
\begin{align}\label{eq:tau_f}\nonumber 
\hat{\tau}&(s(a) \ | \ \mathbf{x}_i)  = \\ 
& f^{s(a) }(\mathbf{x}_i) - f^{s(a)-1}(\mathbf{x}_i), \forall s(a) \in \mathcal{S}\backslash \{0\},
\end{align}
where $f^{s(a)}$ and $f^{s(a)-1}$ predict $E[ y_i (s) \ | \ \mathbf{x}_i] $ and $E[ y_i (s(a)-1) \ | \ \mathbf{x}_i]$, respectively.  To accurately predict the lift-effect $\tau$, it is essential to predict the expected probability of conversion under each ad exposure state appropriately. As a result, the bid price is higher for users with a higher lift-effect. At the same time, the bid price is high for ad-slot with high CTR.

\subsection{Unbiased Lift-effect Prediction}
\label{subsec:attribution}
To obtain a well-performing predictor $f$ for each ad exposure state, it is ideal to directly optimize the following generalization error:
\begin{align} \nonumber \footnotesize
    \mathcal{L}_{ideal} &(f^{s(a)})  \\ &= E_{(\mathbf{x}, y(s(a)))} [ \ell (y (s(a)), f^{s(a)} (\mathbf{x})) ],
\label{eq:ideal_loss}
\end{align}
where $f^{s(a)}$ is a predictor for $ E[y (s(a)) \ | \ \mathbf{x}] $, $\ell$ specifies a loss function such as the mean squared error, $p(x,y(s(a))$ is the joint probability distribution of the entire population, meaning that the population before the ad auction selection or the testing time. We consider optimizing the generalization error defined over the entire population because we apply $f^{s(a)}$ to predict the potential outcome in the testing time. 

 In reality, however, it is impossible to directly optimize Eq.~(\ref{eq:ideal_loss}). This is because we can only utilize a finite size $n_{s(a)}$ of training data $\mathcal{D}_{s(a)} = \{ (\mathbf{x}_i, y_i^{obs}) \ |\ s_i=s(a)\}_{i=1}^{n_{s(a)}} \sim p(x,y  | s=s(a))$ for each ad state and cannot take the expectation of obtaining Eq.~(\ref{eq:ideal_loss}), as we cannot know the exact joint distribution. The conventional solution to this issue is the \textit{empirical risk minimization} (ERM), which optimizes the empirical approximation of Eq.~(\ref{eq:ideal_loss}) as
\begin{align}\nonumber
&f^{s(a)}_{ERM} = \argmin_{f^{s(a)}} \hat{\mathcal{L}}_{ERM} (f^{s(a)}) \\ 
&=  \argmin_{f^{s(a)}}  \frac{1}{n_{s(a)}} \sum_{i \in \mathcal{D}_{s(a)}} \ell (y_i^{obs}, f^{s(a)} (\mathbf{x}_i)).
\label{eq:erm}
\end{align}

The ERM principal works well under the situation of the same train-test distribution, however, the ad impression bias breaks this premise of machine learning. Specifically, the simple empirical approximation of the loss function over $\mathcal{D}_{s(a)}$ has a bias, that is, $ E_{(\mathbf{x}, y(s(a)))} [\hat{\mathcal{L}}_{ERM} (f^{s(a)})] \neq \mathcal{L}_{ideal} (f^{s(a)})$ for a given $f^{s(a)}$.
The bias issue emerges because users assigned to higher bid prices have higher density in the training data than in the test data (i.e., $p(x,y) \neq p (x,y | s=s(a))$).  As a result, the trained predictor $ f^{s(a)}_{ERM}$ may perform poorly in the testing time because it mistakenly overfits the over-represented samples in the training data.

To alleviate this bias issue with ERM in online advertising, we apply the \textit{inverse propensity score} (IPS) estimation technique to debias the estimation of the ideal loss in Eq.~(\ref{eq:ideal_loss}). Our loss function takes the following form,
\begin{align}
    \hat{\mathcal{L}}_{IPS} (f^{s(a)})
     & =  \frac{1}{n} \sum_{i \in \mathcal{D}_{s(a)}} \frac{1}{e_{s(a)} (\mathbf{x}_i)} \ell (y_i^{obs}, f^{s(a)} (\mathbf{x}_i)).
\label{eq:ips_loss}
\end{align}
where $n = \sum_{s(a) \in \mathcal{S}} n_{s(a)}$ is the total number of the training data, and $ e_{s(a)} (\mathbf{x}_i) = P (s_i = s(a)  | \mathbf{x}_i ) $ is the probability that user $i$ is assigned to the ad exposure state $s_i$ called the \textit{propensity score}. A fascinating property of the IPS loss in Eq.~(\ref{eq:ips_loss}) is that it is unbiased for the ideal generalization error as the following proposition states.
\begin{proposition} The IPS loss function in Eq.~(\ref{eq:ips_loss}) is unbiased for the ideal generalization error in Eq.~(\ref{eq:ideal_loss}), that is, for any given $f^{s(a)}$, we have
\begin{align*}
    E_{(\mathbf{x}, \mathbf{y}, s)} [ \hat{\mathcal{L}}_{IPS} (f^{s(a)}) ] = \mathcal{L}_{ideal} (f^{s(a)})
\end{align*}
\begin{proof}
\begin{align*}
    & E_{(\mathbf{x}, \mathbf{y}, s)} [ \hat{\mathcal{L}}_{IPS} (f^{s(a)}) ] \\
    & = E_{(\mathbf{x}, \mathbf{y}, s)} [ \frac{1}{n} \sum_{i \in \mathcal{D}_{s(a)}} \frac{1}{e_{s(a)} (\mathbf{x}_i)} \ell (y_i^{obs}, f^{s(a)} (\mathbf{x}_i)) ] \\
    & = E_{(\mathbf{x}, \mathbf{y}, s)} [ \frac{1}{n} \sum_{i \in \mathcal{D}} \frac{\mathbb{I} \{ s_i = s(a) \}}{e_{s(a)} (\mathbf{x}_i)} \ell (y_i^{obs}, f^{s(a)} (\mathbf{x}_i)) ] \\
    & =  \frac{1}{n} \sum_{i \in \mathcal{D}} E_{\mathbf{x}} [ \frac{E_{s}[ \mathbb{I} \{ s_i = s(a) \} \ | \ \mathbf{x}_i ]}{e_{s(a)} (\mathbf{x}_i)} \cdot \\
    & E_{y (s(a)) } [ \ell (y_i(s(a)), f^{s(a)} (\mathbf{x}_i)) \ | \ \mathbf{x}_i ] ]\\
    & =  \frac{1}{n} \sum_{i \in \mathcal{D}} E_{\mathbf{x}} [ E_{y (s(a)) } [ \ell (y_i(s(a)), f^{s(a)} (\mathbf{x}_i)) \ | \ \mathbf{x}_i ] ]\\
    & = \frac{1}{n} \sum_{i \in \mathcal{D}} E_{(\mathbf{x}, y (s(a) ))} [ \ell (y_i(s(a)), f^{s(a)} (\mathbf{x}_i)) ] \\ 
    & = E_{(\mathbf{x}, y (s(a)))} [ \ell (y (s(a)), f^{s(a)} (\mathbf{x})) ] = \mathcal{L}_{ideal} (f^{s(a)})
\end{align*}
\end{proof}
under standard identification assumptions in causal inference~\cite{imbens2015causal,saito_unbiased_2020,saito2019doubly}.
$\mathcal{D} = \{ (\mathbf{x}_i, y^{obs}_i, s_i )\}_{i=1}^n$ is the size $n$ of the dataset containing all samples.
\end{proposition}

The above proposition suggests that our IPS loss function successfully alleviates the bias issue of ERM and approximates the ideal loss from only observable data. Therefore, to unbiasedly predict the lift-effect under the ad impression bias, we optimize the IPS loss and use the resulting predictors to obtain the final lift-effect prediction as:
\begin{align*}
    \hat{\tau}(s(a) \ | \ \mathbf{x}_i) = f_{IPS}^{s(a)}(\mathbf{x}_i) - f_{IPS}^{s(a)-1}(\mathbf{x}_i), \forall s(a) \in \mathcal{S} \backslash \{0\}.
\end{align*}
where $ f_{IPS}^{s(a)} =  \argmin_{f^{s(a)}} \hat{\mathcal{L}}_{IPS} (f^{s(a)})$ is the IPS loss minimizer.

\section{Online Experiment}
\label{sec: the online experiment}
We conduct a rigorous online experiment using a real-world advertising campaign, the gold standard for evaluation of online systems~\cite{kohavi_tang_xu_2020} to examine the performance of various bidding strategies.

One can consider an evaluation with the offline experiment using past data but it is inappropriate for the DSP. First of all, the realization of ad auctions depends on the complex interplay between competitor DSPs. We need information on the competitors bidding strategy to simulate the real-world RTB environment. Second, the simulation needs the response model of users since we need counterfactual intervention and its response. However, the results inevitably depend on the user model and are not reliable.

On the other hand, an online experiment simply evaluates the performance of bidding systems without further information. We deployed the proposed bidding strategy along with other bidding strategies in a real-world DSP server at CyberAgent, inc, a Japan-based major adtech company. The DSP delivers ads to mobile phone apps through RTB. The DSP follows the CPC billing rule so that it charges the advertiser with a fixed cost per click while the advertiser pursues an increase in foot traffic to the real stores. We deploy five bidders for extensive performance comparison.

\subsection{Setting}
\subsubsection{Comparison Variants}
Table \ref{table:variants} summarizes the five variants deployed in the experiments. 
\begin{table}[htph]\small
\centering
\caption{Comparison Variants for the Experiment.}
\begin{tabular}{m{1cm}m{3cm}m{.9cm}m{.8cm}}
  \toprule
& \multicolumn{1}{c}{description}  & \#users & budget \\ 
  \midrule
\textbf{Baseline}  &  Performance-based& 268,018 & 1.0 \\ 
\textbf{Naive}  & Naive lift-based \`a la \cite{xu2016lift} & 266,917 & 0.1\\ 
\textbf{Unbiased} &  Unbiased lift-based&  267,289 & 0.1  \\ 
\textbf{Noclip}  & Unbiased lift-based w/o clipping \`a la ~\cite{moriwakiUnbiasedLiftbasedBidding2020}& 267,343 & 0.1 \\ 
\textbf{Control}  &  No ad is delivered. & 267,758 & 0.0\\
   \bottomrule
\end{tabular}
\label{table:variants}
\begin{minipage}{\linewidth} 
{\small {\it Note}: Budget is normalized to baseline. Since variant except for baseline is experimental, the assigned budget is small for them. We normalize the key matrices per user and budget for a fair evaluation.  
\par}
\end{minipage}
\end{table}
We split our user base into five groups including the control group. As the bidders except for the baseline are experimental we assign a small fraction of the budget to them. We normalize the key matrices per user and budget for a fair evaluation. The five variants are as follows:
\begin{itemize}[leftmargin=*]
    \item 
    \textbf{Baseline} is a performance-based bidder implemented in the production which bid according to the predicted conversion rate (pCVR, $E[y|{\rm ad}]$) i.e., ${\rm bid}_{t} = \alpha \cdot {\rm CPC} \cdot {\rm pCTR} \cdot {\rm pCVR}$. 
    \item
    \textbf{Unbiased} determines the bid price according to eq.(\ref{eq:bidding}). To make the estimation results stable, we clip the propensity score at top 0.1\%, i.e., $e_{s(a)}(\mathbf{x}_i) = \min(e_{s(a)}(\mathbf{x}_i), \tilde{e}_{s(a)})$ where $\tilde{e}_{s(a)}$ is 99.9 percentile of the propensity score~\cite{leeWeightTrimmingPropensity2011a}. 
    \item
    \textbf{Noclip} is the unbiased lift-based bidder without clipping proposed in ~\cite{moriwakiUnbiasedLiftbasedBidding2020}.
    \item
    \textbf{Naive} is a lift-based bidder without debiasing proposed in ~\cite{xu2016lift}, which is trained by a simple ERM loss (eq.~(\ref{eq:erm})) and substitute the obtained prediction to the bidding model. \item
    \textbf{Control} are not exposed to the ad. Since the experiment measure the foot trafic to real stores even control group can be converted. The number of conversions in this group is the organic conversions (i.e. $E[y|\rm{no \, ad}]$) and used the results to calculate incremental actions. 
\end{itemize}
\subsubsection{Advertising Campaign}
We ran the experiment in an ad campaign by a major department store company that promotes a new lifestyle with new high-quality products. We define visits to real stores as conversion. The ads are delivered to pre-defined Android smartphone users' apps through RTB. 

We measure the number of visits to 70 stores of the company located all over Japan using location data from the audiences' app. The visit is counted each day up to one for each user. Since visits to specific stores located inside or very close to train stations are hard to detect by location data, 10 stores are deleted from the measurements. The experiment ran one week from September 29, 2020.

\subsection{Implementation}
We deployed five variants of bidders in real-world DSP servers. For each bid request, the bidder returns the bid price according to $CPC$, the predicted CTR, and the predicted value of the user $\phi(\mathbf{x}_i, s(a))$. $\phi(\mathbf{x}_i, s(a))$ is performance-based (i.e., the predicted probability of visit) for the baseline and lift-based (i.e., the predicted lift due to the ad) for the lift-based bidders. CTR is predicted using the past CTR of the ad-slots and size of the creative. All the bidders share the same CTR predictor. The input $\mathbf{x}_i$ contains user $i$'s features includes frequency (how many times the user visited the stores) and distance from home to the nearest stores. 
\subsubsection{Architecture}
\label{subsec:architecture}
\begin{figure}[htb]
  \centering
  \includegraphics[width=0.8\linewidth]{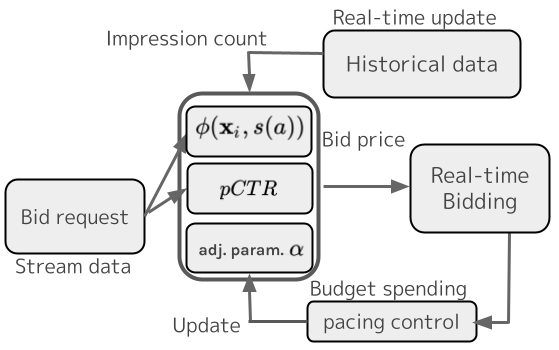}
  \caption{The lift-based bidding system architecture}  \label{fig:system}
\vskip 0.03in
\begin{minipage}{\columnwidth} 
{\small {\it Note}: For each bid request, the DSP server immediately checks the impression count of the same user. The adjustment parameter $\alpha$ is updated every hour based on the budget digestion. The bid price is a combination of $\alpha$, $pCTR$ and $\phi(\cdot)$ (predicted lift-effect or CVR).}
\end{minipage}
\end{figure}

We summarize the whole architecture of our bidding system in Figure~\ref{fig:system}. For each bid request, the corresponding $\phi(\mathbf{x}_i, s(a))$ multiplied by $pCTR$ and $\alpha$ is returned as the bid price. $\phi$ is the predicted lift-effect (CVR for baseline) for a coming impression. The ad impression count is calculated by scanning the history of ad impressions for each user. The entire procedure is scalable and performed within a few milliseconds and does not harm the user experience. 

\subsubsection{Automated Bid Adjustment}
\label{sec:automated bid adjustment}
We use $\alpha$ for budget pacing, and it adjusts the bid price as in Eq.~(\ref{eq:bidding}). Appropriate $\alpha$ is not known a priori, and thus we use an automated bid adjustment algorithm to update it to keep the spending constant. The algorithm increases $\alpha$ when the budget digestion is less than the target and vice versa. 

\subsubsection{Attribution of the Conversion}
Attribution of conversion to each ad impression is a hard task. To circumvent this problem, we follow user-level learning following~\cite{moriwakiUnbiasedLiftbasedBidding2020}. Unlike log-level learning chosen by other studies~\cite{xu2016lift,johnsonCostIncrementalAction2015}, we need neither arbitrary attribution rule nor complex estimation.
\subsubsection{Lift-effect Predictor Training}
\label{subsec: model training}
We train each predictor on the data from past advertising campaigns by the same advertiser. The online ad takes various forms, including ad creative, template, format, and size. Size and ad creative are especially critical. To account for the effect of ad size and creative on target variables, we train specific predictors for each size and creative. As a result, we train $|\mathcal{S}|$ predictors for each ad creative.

The propensity score, $e_{s}(\cdot)$ in Eq.~(\ref{eq:ips_loss}) predict the probability of that the user is exposed to the ad $a$ $s$ times. We use a XGBoost multi-class classifier~\cite{chen_xgboost:_2016} and train it using the whole training data $\{( \mathbf{x}_i, s(a)_i )\}_{i=1}^n$. Hyperparameters are tuned with cross-validation.

The feature vector includes the predicted CVR used for the existing bidder and the number of impressions before the advertising campaign of the training period as well as the other features. These two features improve the accuracy because the higher predicted CVR means a higher bid price and the number of impressions in the training period indicates how easily the ad-slots of the users are obtained. 

For each ad creative, we split users into eight classes by the number of impressions $s_i \in \mathcal{S}=\{0, 1, ,2 ,3, 4, 5-9, 10-19, 20+\}$. The window size of bins increases for a higher number because the distribution of the number of impressions is highly skewed to the right. We train four propensity score predictors for each size and ad creative.

We train outcome predictors for each pair of ad exposure states and ad creative. We use the XGBoost regressor to predict the number of visits to an offline store by weighting each sample by IPS. See Eq.~(\ref{eq:ips_loss}) for the detail. 

\subsection{Results}
\subsubsection{Overall Performance (Table~\ref{table:results})}
CTR is statistically significantly highest for \textbf{unbiased} and almost the same for the other three. This suggests lift-based bidding has no problem in collecting clicks. Mean visits are higher for lift-based bidders (\textbf{naive}, \textbf{noclip} and \textbf{unbiased}) than \textbf{baseline}. This result is consistent with the existing literature~\cite{xu2016lift,moriwakiUnbiasedLiftbasedBidding2020}. One can argue that the baseline is not strong enough. However, the causal effect of advertising is typically very small~\cite{lewisUnfavorableEconomicsMeasuring2015}. The result highlights that the shift from performance-based strategies to lift-based strategies has a significant effect.
\begin{table}[tbhp]\small
\centering
\caption{A/B testing; Overall Performance}
\centering
\begin{tabular}{lrrrr}
  \toprule
 &  \multicolumn{1}{c}{mean CTR }   & \multicolumn{1}{c}{mean visits}    \\ 
  \midrule
  \textbf{baseline} & 1.000 (0.079) & 1.004 (0.016) \\ 
  \textbf{naive} & 1.020 (0.243) & 1.015 (0.015)   \\ 
  \textbf{noclip} & 1.000 (0.229) & \textbf{1.018} (0.016) \\ 
  \textbf{unbiased} & \textbf{1.207} (0.282)   & 1.016 (0.016)  \\ 
 \textbf{control} &   & 1.000 (0.015)  \\
   \bottomrule
\end{tabular}
\label{table:results} 
\vskip 0.03in
\begin{minipage}{.9\linewidth} 
{\small {\it Note}: \textbf{mean CTR}, the mean click-through rate for users with one more impression.; \textbf{mean visits}, the average number of visits during the experiment. All numbers are  divided by the baseline. Standard errors are in parenthesis.
\par}
\end{minipage}
\end{table}

\subsubsection{Cost-efficiency (Table \ref{tab:business})}
Now we further investigate the result by calculating the cost-efficiency of each bidder. The first column of Table~\ref{tab:business} shows the share of inventory cost in CPC charge which represents the share of cost in sales for DSP. Lower is better. The result clearly shows that \textbf{noclip} and \textbf{unbiased} are most profitable for DSP providers. 

Now, we look at metrics for advertisers. First of all, visit lift (the number of visits per user of each bidder minus that of \textbf{control}) is much higher for lift-based bidders and \textbf{noclip} is slightly better than the others.

Finally, we look at CPIA which is calculated as $\frac{\rm CPC \, charge}{\rm incremental \, visits}$. CPIA is of the advertiser's primary interest when checking the campaign results since the metric answer {\it how much the advertiser spent to acquire one more customer's visit?}. Lift-based bidders are much better than performance-based (\textbf{baseline}) and \textbf{noclip} is slightly better than the other two lift-based bidders. This result highlights the cost-efficiency of lift-based bidding strategies for advertisers.

\begin{table}[htb]\small
\centering
\caption{A/B Testing; Cost-efficiency}
\begin{tabular}{lccc}
  \hline
 & \% inv. cost & visit lift & CPIA  \\ 
  \hline
\textbf{baseline} & 1.000 & 1.000 & 1.000 \\ 
  \textbf{naive} & 0.906 & 4.109 & 0.024 \\ 
  \textbf{noclip} & 0.588 & \textbf{4.843} & \textbf{0.021}  \\ 
  \textbf{unbiased} & \textbf{0.547} & 4.393 & 0.023  \\ 
   \hline
\end{tabular}
\vskip 0.03in
\begin{minipage}{.9\linewidth} 
{\small {\it Note}: \textbf{\% inv. cost} is the share of inventory cost to CPC charge; \textbf{visit lift} is the mean visits per user subtracted by that of the control group; \textbf{CPIA} stands for cost per incremental action, the advertiser's cost for each incremental visit. Smaller is better.All numbers are normalized to the baseline. 
}
\end{minipage}
\label{tab:business}
\end{table}

\section{Ablation Study}
We observed compelling results for lift-based bidders. In this section, we disentangle the high performance and cost-efficiency of the lift-based bidders. After comparing the $\phi$ values among the bidders, we investigate their performances in the advertisement auction. This ablation study demonstrates that the mechanism behind the lift-based bidders' high performance is supported by the accuracy and stability in predicting ad slots' value.

\subsection{Comparisons of the Predictions Results}\label{comp_predd_res}
\begin{figure}[htbp]
  \centering
  \includegraphics[width = \columnwidth]{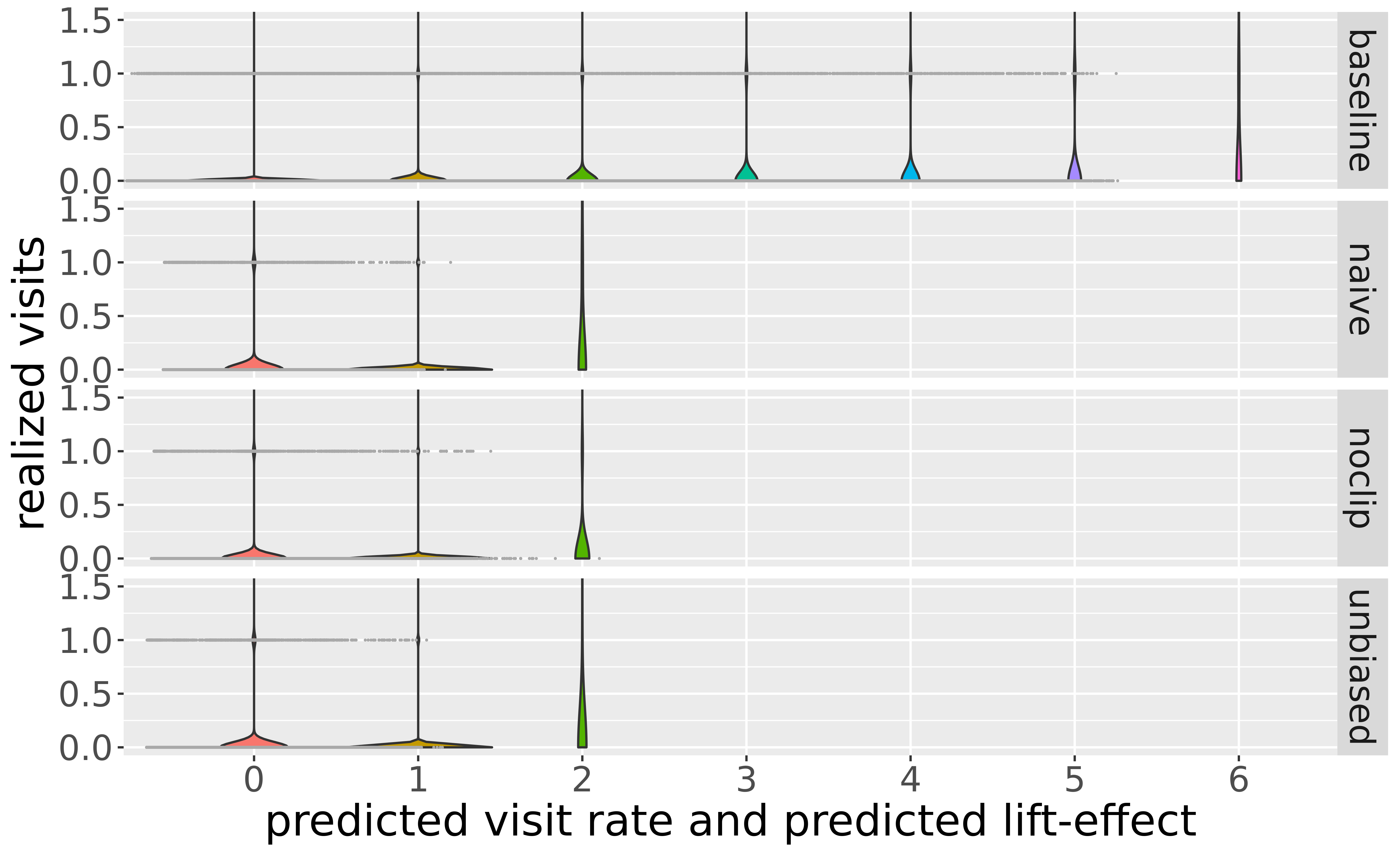}
  \caption{Predicted values ($\phi$) and realized visits}  \label{fig:factor}
  \begin{minipage}{\columnwidth}%
    {\small {\it Note}: The distribution of mean visits of users binned by $\phi$ value. The horizontal axis represents bins on $\phi$ value. 
    }
  \end{minipage}
  \end{figure}

Fig.~\ref{fig:factor} compares the realized number of visits for each bin of $\phi(\cdot)$, $[0, 0.5), [0.5,1.5), \cdots, [5.5,6.5]$. $\phi$ represents pCVR for baseline (performance-based) and lift effect for the others. Since $\phi$ is normalized to have the same mean (eq.~(\ref{eq:phi})), the long tail of the baseline implies that the variance of $\phi(\cdot)$ is much larger for the baseline bidder. It is intuitive because baseline considers not only lift-effect but also random organic visits as attributed visits. In the following study, we will point out this distinction brings competence to the lift-based bidder.
\subsubsection{The inventory cost of ad slots}
The first column of Table~\ref{tab:winrate} shows the average ad inventory cost, which represents how much DSPs pay to buy an impression. The average inventory cost is highest for \textbf{naive}, which is consistent with \cite{xu2016lift}. On the other hand, unbiased lift-based bidders (\textbf{noclip} and \textbf{unbiased}) are less costly than baseline. Even though the unbiased lift-based bidders bid lower prices, the win rates in the second column of Table~\ref{tab:winrate} are almost the same for the four bidders. Unbiased lift-based bidders somehow can save the cost without sacrificing win rates.

  \begin{table}[htbp]\small
\centering
 \caption{Inventory Cost and Win rate}
\begin{tabular}{lcc}
  \hline
 & avg inv. cost & win rate (\%)   \\ 
  \hline
\textbf{baseline} & 1.018  & \textbf{14.1}\\ 
\textbf{naive} & 1.068& 13.6  \\ 
\textbf{unbiased} & 0.835 & 13.3  \\ 
\textbf{noclip} & \textbf{0.765} & 13.4  \\ 
   \hline
\end{tabular}
\vskip 0.03in
\begin{minipage}{\columnwidth} 
{\small {\it Note}: \textbf{avg inv. cost} is the total inventory cost borne by DSP divided by the number of impressions;\textbf{win rate} is \# impressions divided by \# bid requests.The numbers are normalized to the average.
}
\end{minipage}
\label{tab:winrate}
\end{table}

\subsection{Bidding behavior of the lift-based bidders}
To study the source of the cost efficiency of unbiased lift-based bidders, we lastly investigate the bidding behavior of the bidders by comparing with the baseline. 
\subsubsection{Win Prices by Bidders}
\label{sec:bidding}
\begin{figure}[htbp]
  \centering
  \includegraphics[width=\linewidth]{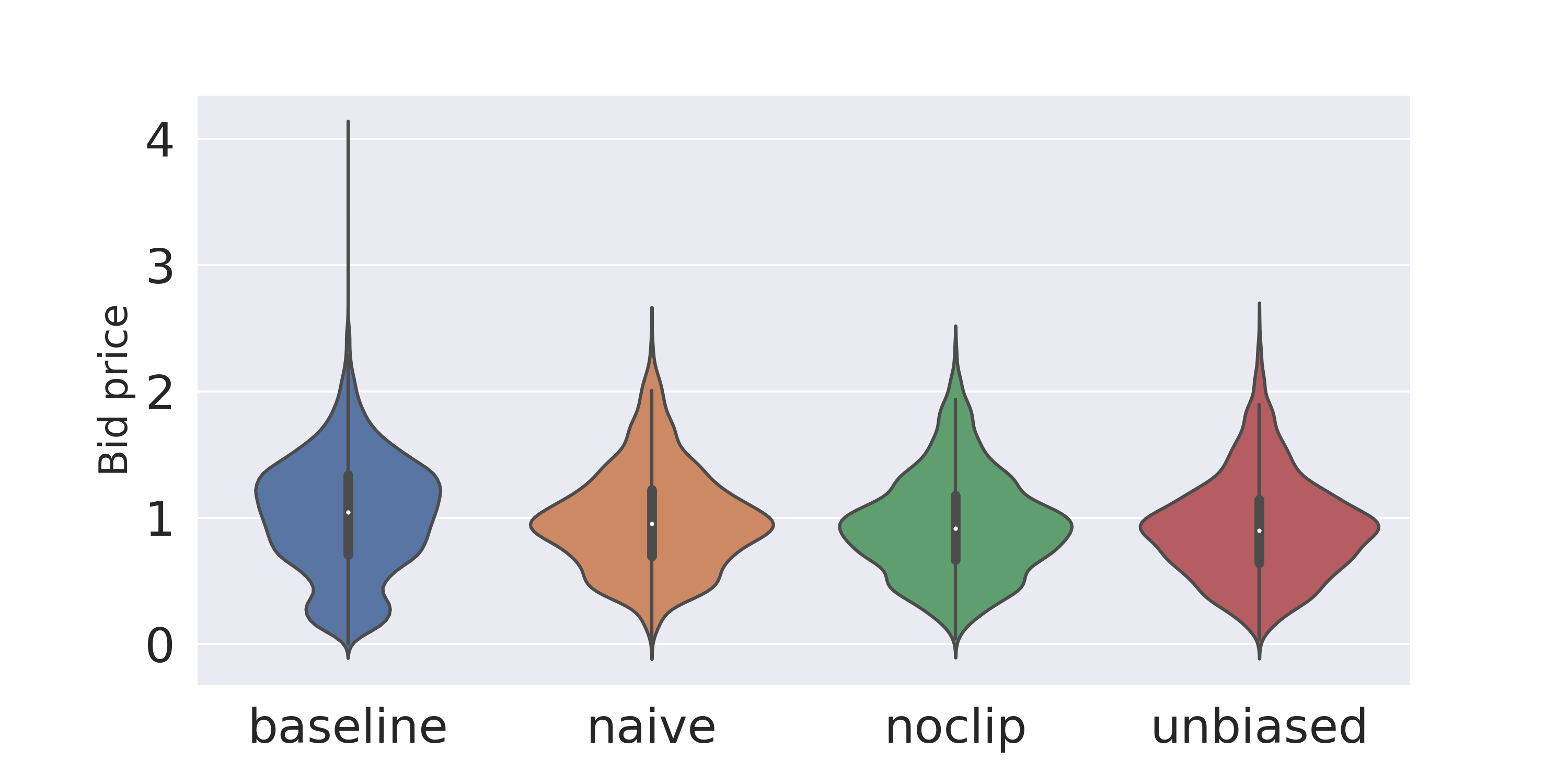}
  \caption{Bid price normalized to the average.}
\label{fig:bid_price}
\vskip 0.03in
\begin{minipage}{\columnwidth} 
{\small {\it Note}: The value of the bidding price history of the wined auctions. The value is on a logarithmic scale and they are normalized to the average of the population. 
\par}
\end{minipage}
\end{figure}

Fig.~\ref{fig:bid_price} describes the distribution of the bid price for ad auction won by the bidders (i.e., win price). The figure represents how the bidders buy impressions. Notice that the actual cost paid by DSPs depends on the type of auctions (i.e., first-price or second-price).

The figure shows that the baseline bids higher prices in general compared to the lift-based bidders. Also, the baseline has a larger variance. As clearly shown, all of the lift-based bidders have a similar distribution of their bidding history. This is consistent with the fact that the baseline tends to predict higher $\phi$ than the lift-based bidders as discussed in Section~\ref{comp_predd_res}. 

\subsubsection{Price Difference between the Bid price and Clearing Price}
What makes unbiased lift-based bidders so economical? To clarify this point, we leverage the fact that a small fraction of ad auctions are run by second-price auctions. In the second price auction, the winner pays the second-highest bid price as the {\it clearing price}. The differences between the winner's bid price and clearing price represent the gap between the ad-slots values predicted by the winner DSP and that by the competitor DSPs. A large gap suggests that the winner would over-pay in the first-price auctions which share most of the auctions.  

Table~\ref{table:pricedif} compares the average price difference for each bidder. Lift-based bidders, especially our favored unbiased lift-based bidders have a smaller gap between bid prices and clearing prices. This suggests that it saves money by bidding close to clearing prices in the first-price auctions. In other words, the baseline overestimates the value of the ad-slots as it ignores the organic visit and the naive lift-based bidder ignores the bias in the training data. Interestingly, without clipping for propensity score the unbiased lift-effect bidder has a larger price difference than the naive lift-based bidder. 

In sum, our ablation study provides our unbiased lift-based bidding system
{\setlength{\leftmargini}{10pt}  
\begin{itemize}
    \item the baseline is more costly than the unbiased lift-based bidders
    \item the lift-based bidder saves inventory cost by bidding close to the clearing price in the first price auctions
    \item clipping further reduces inventory cost
\end{itemize}}

  \begin{table}[hbpt]\small
\centering
 \caption{Average Price difference from the clearing price}
\begin{tabular}{lrrr}
  \hline
 & price diff & gap from baseline& p-val \\ 
  \hline
\textbf{baseline}  & 0.9774 & 0.0 (0.0)  & N/A \\
\textbf{naive}  & 0.8756 & -0.1018 (0.052) & 0.049 \\ 
\textbf{unbiased}  & \textbf{0.8485} & \textbf{-0.1289} (0.063)& 0.040 \\ 
\textbf{noclip}  & 0.9027 & -0.0747 (0.056)  & 0.056 \\ 
   \hline
\end{tabular}
\vskip 0.03in
\begin{minipage}{\columnwidth} 
{\small {\it Note}: The price difference is normalized to the average. The third column shows the estimated difference from the baseline. Standard errors are in parenthesis.
}
\end{minipage}
\label{table:pricedif}
\end{table}
\section{Conclusion}
In this study, we address the practical difficulty in introducing a lift-based bidding in the real world.
We combined a CTR predictor with lift-effect predictors to pursue high CTR which is good for DSPs and low CPIA which is good for advertisers. Then we embedded the various lift effect predictors proposed in the existing research into the bidding systems and compared them along with the performance-based system in an online experiment. As a result, the lift-based bidding system achieved high cost-efficiency without hurting DSP's profit.

Our ablation study shows that (i)the lift-based bidding saves inventory cost by bidding at appropriate prices and (ii)debiasing and clipping further improve cost-efficiency. The online advertising community can find a way to adopt lift-based advertising which fulfills advertisers' true goals (increase in sales) and ultimately contributes to the healthy growth of the industry.
\bibliographystyle{IEEEtran}
\bibliography{ref}

\begin{thebibliography}{10}
\providecommand{\url}[1]{#1}
\csname url@samestyle\endcsname
\providecommand{\newblock}{\relax}
\providecommand{\bibinfo}[2]{#2}
\providecommand{\BIBentrySTDinterwordspacing}{\spaceskip=0pt\relax}
\providecommand{\BIBentryALTinterwordstretchfactor}{4}
\providecommand{\BIBentryALTinterwordspacing}{\spaceskip=\fontdimen2\font plus
\BIBentryALTinterwordstretchfactor\fontdimen3\font minus
  \fontdimen4\font\relax}
\providecommand{\BIBforeignlanguage}[2]{{%
\expandafter\ifx\csname l@#1\endcsname\relax
\typeout{** WARNING: IEEEtran.bst: No hyphenation pattern has been}%
\typeout{** loaded for the language `#1'. Using the pattern for}%
\typeout{** the default language instead.}%
\else
\language=\csname l@#1\endcsname
\fi
#2}}
\providecommand{\BIBdecl}{\relax}
\BIBdecl

\bibitem{USProgrammaticDigital}
eMarketer, ``{{US Programmatic Digital Display Ad Spending}},''
  https://www.emarketer.com/content/us-programmatic-digital-display-ad-spending,
  2019.

\bibitem{lewis2018incrementality}
\BIBentryALTinterwordspacing
R.~A. Lewis and J.~Wong, ``\BIBforeignlanguage{en}{Incrementality {Bidding} \&
  {Attribution}},'' Social Science Research Network, Rochester, NY, {SSRN}
  {Scholarly} {Paper} ID 3129350, Feb. 2018. [Online]. Available:
  \url{https://papers.ssrn.com/abstract=3129350}
\BIBentrySTDinterwordspacing

\bibitem{xu2016lift}
J.~{Xu}, X.~{Shao}, J.~{Ma}, K.~chih {Lee}, H.~{Qi}, and Q.~{Lu}, ``Lift-based
  bidding in ad selection,'' in \emph{AAAI’16: Proceedings of the Thirtieth
  AAAI Conference on Artificial Intelligence}, 2016, pp. 651--657.

\bibitem{moriwakiUnbiasedLiftbasedBidding2020}
D.~Moriwaki, Y.~Hayakawa, I.~Munemasa, Y.~Saito, and A.~Matsui, ``Unbiased
  {{Lift}}-based {{Bidding System}},'' \emph{Proceedings of the AdKDD '20},
  Aug. 2020.

\bibitem{agarwalSpatiotemporalModelsEstimating2009a}
D.~Agarwal, B.-C. Chen, and P.~Elango,
  ``\BIBforeignlanguage{en}{Spatio-temporal models for estimating click-through
  rate},'' in \emph{\BIBforeignlanguage{en}{Proceedings of the 18th
  International Conference on {{World}} Wide Web - {{WWW}} '09}}.\hskip 1em
  plus 0.5em minus 0.4em\relax {Madrid, Spain}: {ACM Press}, 2009, p.~21.

\bibitem{maUserFatigueOnline2016a}
H.~Ma, X.~Liu, and Z.~Shen, ``User {{Fatigue}} in {{Online News
  Recommendation}},'' in \emph{Proceedings of the 25th {{International
  Conference}} on {{World Wide Web}}}, ser. {{WWW}} '16.\hskip 1em plus 0.5em
  minus 0.4em\relax {Republic and Canton of Geneva, Switzerland}:
  {International World Wide Web Conferences Steering Committee}, 2016, pp.
  1363--1372.

\bibitem{moriwaki_fatigue-aware_2020}
\BIBentryALTinterwordspacing
D.~Moriwaki, K.~Fujita, S.~Yasui, and T.~Hoshino, ``Fatigue-{Aware} {Ad}
  {Creative} {Selection},'' \emph{arXiv:1908.08936 [cs, stat]}, Jan. 2020,
  arXiv: 1908.08936. [Online]. Available: \url{http://arxiv.org/abs/1908.08936}
\BIBentrySTDinterwordspacing

\bibitem{criteoIncrementalitySimpleQuestion}
Criteo. (2020) Incrementality: A simple question commanding subtle answers | by
  {{Pl Mrcy}} | {{Criteo R}}\&{{D Blog}} | {{Medium}}.

\bibitem{barajas_online_2021}
J.~Barajas, N.~Bhamidipati, and J.~G. Shanahan, ``Online {Advertising}
  {Incrementality} {Testing} {And} {Experimentation}: {Industry} {Practical}
  {Lessons},'' in \emph{Proceedings of KDD'21}.\hskip 1em plus 0.5em minus
  0.4em\relax New York, NY, USA: Association for Computing Machinery, Aug.
  2021, pp. 4027--4028.

\bibitem{beeswaxMeasuringIncrementalityDigital}
\BIBentryALTinterwordspacing
Beeswax. (2020) Measuring {{Incrementality In Digital Media}}. [Online].
  Available: \url{https://blog.beeswax.com/incrementality}
\BIBentrySTDinterwordspacing

\bibitem{yuan2013real}
S.~Yuan, J.~Wang, and X.~Zhao, ``Real-time bidding for online advertising:
  measurement and analysis,'' in \emph{Proceedings of the Seventh International
  Workshop on Data Mining for Online Advertising}, 2013, pp. 1--8.

\bibitem{wang2015real}
J.~Wang and S.~Yuan, ``Real-time bidding: A new frontier of computational
  advertising research,'' in \emph{Proceedings of the Eighth ACM International
  Conference on Web Search and Data Mining}, 2015, pp. 415--416.

\bibitem{wang2016display}
J.~Wang, W.~Zhang, and S.~Yuan, ``Display advertising with real-time bidding
  (rtb) and behavioural targeting,'' \emph{arXiv preprint arXiv:1610.03013},
  2016.

\bibitem{balseiro2015repeated}
S.~R. Balseiro, O.~Besbes, and G.~Y. Weintraub, ``Repeated auctions with
  budgets in ad exchanges: Approximations and design,'' \emph{Management
  Science}, vol.~61, no.~4, pp. 864--884, 2015.

\bibitem{balseiro2017budget}
S.~Balseiro, A.~Kim, M.~Mahdian, and V.~Mirrokni, ``Budget management
  strategies in repeated auctions,'' in \emph{WWW}, 2017, pp. 15--23.

\bibitem{balseiro2019learning}
S.~R. Balseiro and Y.~Gur, ``Learning in repeated auctions with budgets: Regret
  minimization and equilibrium,'' \emph{Management Science}, vol.~65, no.~9,
  pp. 3952--3968, 2019.

\bibitem{conitzer2018pacing}
V.~Conitzer, C.~Kroer, D.~Panigrahi, O.~Schrijvers, E.~Sodomka, N.~E.
  Stier-Moses, and C.~Wilkens, ``Pacing equilibrium in first-price auction
  markets,'' in \emph{Proceedings of the 2019 ACM Conference on Economics and
  Computation}, ser. EC ’19.\hskip 1em plus 0.5em minus 0.4em\relax New York,
  NY, USA: Association for Computing Machinery, 2019, p. 587.

\bibitem{panBidShadingWinRate2020}
S.~Pan, B.~Kitts, T.~Zhou, H.~He, B.~Shetty, A.~Flores, D.~Gligorijevic,
  J.~Pan, T.~Mao, S.~Gultekin, and J.~Zhang, ``Bid {Shading} by {Win}-{Rate}
  {Estimation} and {Surplus} {Maximization},'' in \emph{Proceedings of
  {ADKDD}'20}, Sep. 2020.

\bibitem{wu2018budget}
D.~Wu, X.~Chen, X.~Yang, H.~Wang, Q.~Tan, X.~Zhang, J.~Xu, and K.~Gai, ``Budget
  constrained bidding by model-free reinforcement learning in display
  advertising,'' in \emph{CIKM}, 2018, pp. 1443--1451.

\bibitem{cai2017real}
H.~Cai, K.~Ren, W.~Zhang, K.~Malialis, J.~Wang, Y.~Yu, and D.~Guo, ``Real-time
  bidding by reinforcement learning in display advertising,'' in
  \emph{Proceedings of WSDM'17}, 2017, pp. 661--670.

\bibitem{yang2019bid}
X.~Yang, Y.~Li, H.~Wang, D.~Wu, Q.~Tan, J.~Xu, and K.~Gai, ``Bid optimization
  by multivariable control in display advertising,'' in \emph{Proceedings of
  the 25th ACM SIGKDD International Conference on Knowledge Discovery \& Data
  Mining}, 2019, pp. 1966--1974.

\bibitem{maehara2018optimal}
T.~Maehara, A.~Narita, J.~Baba, and T.~Kawabata, ``Optimal bidding strategy for
  brand advertising.'' in \emph{IJCAI}, 2018, pp. 424--432.

\bibitem{zhang2016feedback}
W.~Zhang, Y.~Rong, J.~Wang, T.~Zhu, and X.~Wang, ``Feedback control of
  real-time display advertising,'' in \emph{Proceedings of the Ninth ACM
  International Conference on Web Search and Data Mining}, 2016, pp. 407--416.

\bibitem{lewis2014online}
R.~A. Lewis and D.~H. Reiley, ``Online ads and offline sales: measuring the
  effect of retail advertising via a controlled experiment on yahoo!''
  \emph{Quantitative Marketing and Economics}, vol.~12, no.~3, pp. 235--266,
  2014.

\bibitem{cheng2010personalized}
H.~Cheng and E.~Cant{\'u}-Paz, ``Personalized click prediction in sponsored
  search,'' in \emph{Proceedings of the third ACM international conference on
  Web search and data mining}, 2010, pp. 351--360.

\bibitem{zhu2010novel}
Z.~A. Zhu, W.~Chen, T.~Minka, C.~Zhu, and Z.~Chen, ``A novel click model and
  its applications to online advertising,'' in \emph{WSDM'2010}, 2010, pp.
  321--330.

\bibitem{ren2016user}
K.~Ren, W.~Zhang, Y.~Rong, H.~Zhang, Y.~Yu, and J.~Wang, ``User {Response}
  {Learning} for {Directly} {Optimizing} {Campaign} {Performance} in {Display}
  {Advertising},'' \emph{CIKM}, 2016.

\bibitem{qu2016product}
Y.~Qu, H.~Cai, K.~Ren, W.~Zhang, Y.~Yu, Y.~Wen, and J.~Wang, ``Product-based
  neural networks for user response prediction,'' in \emph{ICDM}.\hskip 1em
  plus 0.5em minus 0.4em\relax IEEE, 2016, pp. 1149--1154.

\bibitem{rosales2012post}
R.~Rosales, H.~Cheng, and E.~Manavoglu, ``Post-click conversion modeling and
  analysis for non-guaranteed delivery display advertising,'' in \emph{WSDM},
  2012, pp. 293--302.

\bibitem{yeo2017predicting}
J.~Yeo, S.~Kim, E.~Koh, S.-w. Hwang, and N.~Lipka, ``Predicting online purchase
  conversion for retargeting,'' in \emph{Proceedings of WSDM'17}, 2017, pp.
  591--600.

\bibitem{radcliffe2011real}
N.~J. Radcliffe and P.~D. Surry, ``Real-world uplift modelling with
  significance-based uplift trees,'' \emph{Portrait Technical Report TR-2011-1,
  Stochastic Solutions}, 2011.

\bibitem{jaskowski2012uplift}
M.~Jaskowski and S.~Jaroszewicz, ``Uplift modeling for clinical trial data,''
  in \emph{ICML Workshop on Clinical Data Analysis}, 2012.

\bibitem{rzepakowski2012decision}
P.~Rzepakowski and S.~Jaroszewicz, ``Decision trees for uplift modeling with
  single and multiple treatments,'' \emph{Knowledge and Information Systems},
  vol.~32, no.~2, pp. 303--327, 2012.

\bibitem{zaniewicz2013support}
L.~Zaniewicz and S.~Jaroszewicz, ``Support vector machines for uplift
  modeling,'' in \emph{ICDMW ’13: Proceedings of the 2013 IEEE 13th
  International Conference on Data Mining Workshops}.\hskip 1em plus 0.5em
  minus 0.4em\relax IEEE, 2013, pp. 131--138.

\bibitem{diemertLargeScaleBenchmark2018}
E.~Diemert, A.~Betlei, C.~Renaudin, and M.-R. Amini,
  ``\BIBforeignlanguage{en}{A {Large} {Scale} {Benchmark} for {Uplift}
  {Modeling}},'' in \emph{\BIBforeignlanguage{en}{AdKDD'18}}, London United
  Kingdom, 2018, p.~6.

\bibitem{kawanaka2019}
S.~Kawanaka and D.~Moriwaki, ``Uplift modeling for location-based online
  advertising,'' in \emph{Proceedings of the 3rd ACM SIGSPATIAL International
  Workshop on Location-Based Recommendations, Geosocial Networks and
  Geoadvertising}, 2019, pp. 1--4.

\bibitem{saito2019doubly}
Y.~Saito, H.~Sakata, and K.~Nakata, ``Doubly robust prediction and evaluation
  methods improve uplift modeling for observational data,'' in
  \emph{Proceedings of the 2019 SIAM International Conference on Data
  Mining}.\hskip 1em plus 0.5em minus 0.4em\relax SIAM, 2019, pp. 468--476.

\bibitem{saito2020cost}
------, ``Cost-effective and stable policy optimization algorithm for uplift
  modeling with multiple treatments,'' in \emph{Proceedings of the 2020 SIAM
  International Conference on Data Mining}.\hskip 1em plus 0.5em minus
  0.4em\relax SIAM, 2020, pp. 406--414.

\bibitem{johnsonCostIncrementalAction2015}
G.~A. Johnson and R.~A. Lewis, ``\BIBforeignlanguage{en}{Cost {{Per Incremental
  Action}}: {{Efficient Pricing}} of {{Advertising}}},''
  \emph{\BIBforeignlanguage{en}{SSRN Electronic Journal}}, 2015.

\bibitem{bottou_counterfactual_2013}
L.~Bottou, J.~Peters, J.~Quiñonero-Candela, D.~X. Charles, D.~M. Chickering,
  E.~Portugaly, D.~Ray, P.~Simard, and E.~Snelson, ``Counterfactual {Reasoning}
  and {Learning} {Systems}: {The} {Example} of {Computational} {Advertising},''
  \emph{Journal of Machine Learning Research}, vol.~14, no.~65, pp. 3207--3260,
  2013.

\bibitem{joachims_unbiased_2018}
T.~Joachims, A.~Swaminathan, and T.~Schnabel,
  ``\BIBforeignlanguage{en}{Unbiased {Learning}-to-{Rank} with {Biased}
  {Feedback}},'' in \emph{\BIBforeignlanguage{en}{Proceedings of the {Tenth}
  {ACM} {International} {Conference} on {Web} {Search} and {Data}
  {Mining}}}.\hskip 1em plus 0.5em minus 0.4em\relax Cambridge United Kingdom:
  ACM, Feb. 2017, pp. 781--789.

\bibitem{zhang2016bid}
W.~{Zhang}, T.~{Zhou}, J.~{Wang}, and J.~{Xu}, ``Bid-aware gradient descent for
  unbiased learning with censored data in display advertising,'' in \emph{KDD
  ’16}, 2016, pp. 665--674.

\bibitem{bompaire_causal_2021}
M.~Bompaire, A.~Gilotte, and B.~Heymann, ``\BIBforeignlanguage{en}{Causal
  {Models} for {Real} {Time} {Bidding} with {Repeated} {User}
  {Interactions}},'' in \emph{\BIBforeignlanguage{en}{Proceedings of the 27th
  {ACM} {SIGKDD}}}.\hskip 1em plus 0.5em minus 0.4em\relax Virtual Event
  Singapore: ACM, Aug. 2021, pp. 75--85.

\bibitem{imbens2015causal}
G.~W. Imbens and D.~B. Rubin, \emph{Causal inference in statistics, social, and
  biomedical sciences}.\hskip 1em plus 0.5em minus 0.4em\relax Cambridge
  University Press, 2015.

\bibitem{lewis_worn-out_2015}
R.~A. Lewis, ``\BIBforeignlanguage{en}{Worn-{Out} or {Just} {Getting}
  {Started}? {The} {Impact} of {Frequency} in {Online} {Display}
  {Advertising}},'' Boston, Massachusetts, USA, Jan. 2015.

\bibitem{saito_unbiased_2020}
Y.~{Saito}, S.~{Yaginuma}, Y.~{Nishino}, H.~{Sakata}, and K.~{Nakata},
  ``Unbiased recommender learning from missing-not-at-random implicit
  feedback,'' in \emph{WSDM 2020: The 13th ACM International Conference on Web
  Search and Data Mining}, 2020, pp. 501--509.

\bibitem{kohavi_tang_xu_2020}
R.~Kohavi, D.~Tang, and Y.~Xu, \emph{Trustworthy Online Controlled Experiments:
  A Practical Guide to A/B Testing}.\hskip 1em plus 0.5em minus 0.4em\relax
  Cambridge University Press, 2020.

\bibitem{leeWeightTrimmingPropensity2011a}
B.~K. Lee, J.~Lessler, and E.~A. Stuart, ``Weight {{Trimming}} and {{Propensity
  Score Weighting}},'' \emph{PLoS ONE}, vol.~6, no.~3, Mar. 2011.

\bibitem{chen_xgboost:_2016}
T.~{Chen} and C.~{Guestrin}, ``Xgboost: A scalable tree boosting system,'' in
  \emph{KDD ’16: Proceedings of the 22nd ACM SIGKDD International Conference
  on Knowledge Discovery and Data Mining}, 2016, pp. 785--794.

\bibitem{lewisUnfavorableEconomicsMeasuring2015}
R.~A. Lewis and J.~M. Rao, ``\BIBforeignlanguage{en}{The {{Unfavorable
  Economics}} of {{Measuring}} the {{Returns}} to {{Advertising}}},''
  \emph{\BIBforeignlanguage{en}{The Quarterly Journal of Economics}}, vol. 130,
  no.~4, pp. 1941--1973, Nov. 2015.

\end{thebibliography}

\end{document}